%% file: main.tex
\documentclass{article}
\usepackage{spconf,amsmath,amsfonts,amssymb,graphicx}
\usepackage[acronym,toc,shortcuts,nonumberlist]{glossaries}
\usepackage{multirow}
\usepackage{subcaption}
\usepackage{color}
\usepackage{csquotes}  %
\usepackage[table]{xcolor}
\usepackage{xspace}
\usepackage{hyperref}
\usepackage{cleveref}
\usepackage{siunitx}
\usepackage[absolute,showboxes]{textpos}
\usepackage{pifont}

\input{acronyms}
\input{commands}

\newif\iffinal
\finaltrue

\title{Combining TF-GridNet and Mixture Encoder for\\Continuous Speech Separation for Meeting Transcription}
\iffinal
\name{
\begin{tabular}{@{}c@{}}
Peter Vieting$^{1}$, Simon Berger$^{1,2}$, Thilo von Neumann$^{3}$, Christoph Boeddeker$^{3}$,\\Ralf Schl\"uter$^{1,2}$, Reinhold Haeb-Umbach$^{3}$
\end{tabular}
}
\address{
$^1$Machine Learning and Human Language Technology Group, RWTH Aachen University, Germany\\
$^2$AppTek GmbH, Germany\quad
$^3$Paderborn University, Germany\\
\textit{\{vieting,sberger,schlueter\}@hltpr.rwth-aachen.de, \{vonneumann,boeddeker,haeb\}@nt.upb.de}
}
\else
\name{BLIND}
\address{BLIND}
\fi

\begin{document}
\TPMargin{5pt}
\begin{textblock*}{1.05\textwidth}(1.4cm, 26cm)
    \noindent
    \small
    © 2024 IEEE. Personal use of this material is permitted. Permission from IEEE must be obtained for all other uses, in any current or future media, including reprinting/republishing this material for advertising or promotional purposes, creating new collective works, for resale or redistribution to servers or lists, or reuse of any copyrighted component of this work in other works.
\end{textblock*}
\maketitle
\begin{abstract}
Many real-life applications of \gls{ASR} require processing of overlapped speech.
A common method involves first separating the speech into overlap-free streams on which \gls{ASR} is performed.
Recently, \tfgridnet has shown impressive performance in speech separation in real reverberant conditions.
Furthermore, a mixture encoder was proposed that leverages the mixed speech to mitigate the effect of separation artifacts.
In this work, we extended the mixture encoder from a static two-speaker scenario to a natural meeting context featuring an arbitrary number of speakers and varying degrees of overlap.
We further demonstrate its limits by the integration with separators of varying strength including \tfgridnet.
Our experiments result in a new state-of-the-art performance on \libricss using a single microphone.
They show that \tfgridnet largely closes the gap between previous methods and oracle separation independent of mixture encoding.
We further investigate the remaining potential for improvement.
\end{abstract}
\glsresetall
\section{Introduction}
The transcription of meetings is an important application scenario for \gls{ASR}.
However, it is inherently difficult for multiple reasons, e.g., the spontaneous nature of the speech, reverberation in the room or background noise \cite{yu2022m2met}.
Another vital aspect is the appropriate handling of speaker overlap.
Past work on speech separation has largely focused on fully overlapping speech of only two speakers \cite{yu2017pit,luo2019convtasnet,drude2019smswsj}.
Recently, there has been a shift in attention towards \gls{CSS} for realistic meeting scenarios \cite{yoshioka2018CSS,chen2020libricss,Raj2021}.

The methods for multi-speaker \gls{ASR} can generally be divided into modular and end-to-end approaches.
While end-to-end approaches handle separation and recognition jointly and just take a single, global decision, modular approaches may utilize dedicated separation front-ends.
Among those separation front-ends, \tfgridnet \cite{wang2023tfgridnet_transactions, wang2023tfgridnet_icassp} has recently shown great performance in separating overlapped and noisy speech in the presence of reverberation.
Furthermore, a mixture encoder has been proposed to improve the modular approach by providing information about the original speech mixture in addition to the single separated speaker \cite{berger2023interspeech}.
However, the stronger the separator, the less improvement can be expected by this method.
Therefore, we test its limits by combining it with separators of different strengths.

The main contributions of this work are
\begin{itemize}
    \item the extension of the mixture encoder to a meeting scenario with an arbitrary number of speakers and varying degrees of overlap using the \gls{CSS} approach,
    \item state-of-the-art results on \libricss using a single microphone achieved by the combination with the powerful \tfgridnet separation front-end,
    \item studying the effect of exploiting mixture information for acoustic modeling in combination with separators of different strengths as well as
    \item investigating the remaining difference to oracle performance with a novel detailed analysis of the effect of imperfect separation on the speech recognition search space by computing coincidence rates between the reference speaker and the cross channel, which provides additional insights into the capabilities of TF-GridNet.
\end{itemize}

\section{Related work}
Transcribing meetings requires a more sophisticated handling of overlapping speech than classical two-speaker speech separation models.
In \gls{CSS}, continuous audio streams are separated into multiple non-overlapped streams \cite{chen2020libricss}.
Under the assumption that not more than two speakers are active at a time, a separator can output two channels and diarization can be used to track the individual speakers in these channels \cite{yoshioka2018CSS}.

\tfgridnet \cite{wang2023tfgridnet_transactions} recently excelled in speech separation under reverberant conditions.
Its application to a meeting scenario is discussed in \cite{vonneumann2023_Meeting}.
The mixture encoder, which enhances a modular approach for multi-speaker \gls{ASR} by utilizing additional information about the mixture signal in the \gls{AM}, has previously been employed on fully overlapped speech \cite{berger2023interspeech} in combination with a \gls{BLSTM} separator.
Unlike in end-to-end approaches, the speech separation front-end can easily be replaced due to the model's modularity.
In this work, we further combine \tfgridnet and the mixture encoder for meeting transcription.

\input{figures/architecture}

\section{Methods}
\vspace{-2mm}

\label{sec:methods}
The baseline architecture of the modular approach is depicted in \Cref{fig:architecture_base}.
During separation, the speech mixture $\mathbf{x}$ is separated in two channels $\hat{\mathbf{x}}^0$ and $\hat{\mathbf{x}}^1$ which contain estimates of the non-overlapped speech.
After that, a simple energy-based \gls{VAD} mechanism is applied to determine the speech segments in each output channel separately.
These segments are input to a single encoder called the separation encoder here.
After applying upsampling via transposed convolution, a linear mapping and a softmax, we obtain the frame-wise label posteriors $p(\mathbf{y}_t^s|\hat{\mathbf{x}}^s)$ for channel $s\in\{0,1\}$.

Similar to \cite{berger2023interspeech}, a mixture encoder is added in \Cref{fig:architecture_mixenc}.
It receives the original speech mixture $\mathbf{x}$ with possibly overlapping speech as input.
The segment boundaries are taken from the single-channel \gls{VAD}.
Note that feature extraction and downsampling are considered to be part of the separation and mixture encoder blocks in \Cref{fig:architecture}.
The separation and mixture encoders' outputs are concatenated and further processed by a linear projection and another encoder, i.e., the \gls{MAS} encoder.
Unlike \cite{berger2023interspeech}, no combination layer is used here and the channels are handled separately.

\vspace{-3mm}
\section{Experimental setup}
\vspace{-3mm}
\subsection{Data}
\vspace{-1mm}
We use the \libricss \cite{chen2020libricss} dataset to evaluate our models on a task with realistic room acoustics and partially overlapped speech.
LibriCSS is derived from the LibriSpeech \cite{panayotov2015librispeech} \textit{test-clean} dataset and features re-recordings of utterances with different overlap ratios set in a meeting room environment.
It is a popular dataset for evaluating multi-speaker \gls{ASR} systems due to its real noise and room acoustics paired with controlled overlap ratios in different subsets \cite{chen2020libricss,kanda2022tsot,boeddeker2023tssep}.

As LibriCSS offers only evaluation data, we use the original LibriSpeech corpus to simulate training data.
This involves producing spatialized and mixed audios, mirroring the process used for the SMS-WSJ dataset \cite{drude2019smswsj}.
Dynamic mixing \cite{zeghidour2021wavesplit,cord2022mms} further helps to increase the amount of generated training data.
The speech separators are trained exclusively on this simulated dataset.

For the \gls{AM}, we first train a baseline single-speaker model on the LibriSpeech corpus.
This initial model serves as the base for initialization as described in \Cref{sec:acoustic_model}.
After initialization, the \gls{AM} is fine-tuned using separated audio tracks which are obtained by applying the previously trained separators to the simulated training data.
During this \gls{AM} fine-tuning phase, we incorporate oracle information for \gls{VAD}.
The simulated \gls{AM} training data amounts to about \SI{1000}{\hour} after \gls{VAD} application.
The standard LibriSpeech text corpus is used for \gls{LM} training.

Since the focus of this work is on single-microphone techniques, only the first microphone of the \libricss recordings is used during evaluation.
As suggested in \cite{chen2020libricss}, we use \textit{Session0} as a dev set to tune the recognition hyperparameters and present results on the test set, i.e., the remaining sessions excluding \textit{Session0}.

\vspace{-3mm}
\subsection{Speech separation model}
\vspace{-1mm}
\label{sec:speech_separator}
The separator training follows the \gls{PIT} paradigm \cite{kolbaek2017multitalker} with a \gls{SA-SDR} loss \cite{neumann2022sasdr} in the time-domain on fully overlapping data and single-speaker utterances.
At test time, the \gls{CSS} idea from \cite{yoshioka2018CSS} is used:
The separator is applied on a sliding window on the observation
with a size of 4s and a shift of 3s so that an overlap of 1s is present between adjacent windows.
This window size is small enough to allow the assumption that at maximum two speakers are active within a single window.
Thus, we obtain two outputs with overlap-free speech.
To resolve the permutation problem between adjacent segments, which arises because the separator can order its outputs arbitrarily, we determine the permutation with minimal \gls{MSE} on the overlapping parts between the segments.
This way, the model can handle an arbitrary number of speakers.

As separators, we train a 3-layer \gls{BLSTM} \cite{kolbaek2017multitalker} as well as \tfgridnet \cite{wang2023tfgridnet_transactions}.
The \gls{BLSTM} architecture is simple, yet competitive with the best models on reverberant data \cite{cord2022monaural}.
Only recently, \tfgridnet advanced the state-of-the-art performance by a large margin for source separation.
Additionally, \tfgridnet was published as part of ESPnet \cite{watanabe2018espnet} and is therefore well reproducible.
Note that the increased performance comes at the price of an increase in execution time by a factor of roughly 40 compared to the \gls{BLSTM}.

\vspace{-3mm}
\subsection{Acoustic model training}
\vspace{-1mm}
\label{sec:acoustic_model}
A hybrid hidden Markov model is used to perform \gls{ASR}.
To generate an alignment between the extracted Gammatone features and the 12001 generalized triphone state labels obtained by a \gls{CART}, we employ a Gaussian mixture hidden Markov model baseline similar to \cite{luescher2019librispeech}.
Note that the alignments are taken from LibriSpeech, i.e., they are computed on the clean single-speaker utterances by a model trained on the regular LibriSpeech data.
They are then edited and padded with silence to fit the simulated data.
The model is trained with a \gls{fCE} loss on single-speaker speech which was produced by the speech separation model.

Unlike \cite{luescher2019librispeech}, we use a Conformer-based \gls{AM} \cite{gulati2020conformer}.
In particular, the neural architecture in the baseline uses a convolutional front-end with a sub-sampling factor of 3 followed by a linear layer and 12 512-dimensional Conformer blocks with 8 attention heads.
Before the softmax over the output labels, we apply a transposed convolution to upsample by a factor of 3 again.
SpecAugment \cite{park2019specaug}, dropout and an additional auxiliary loss with the same targets after the $\text{6}^{\text{th}}$ Conformer block are used for regularization.

The architecture with the mixture encoder is based on \cite{berger2023interspeech}.
Unlike the baseline, it does not only take the separated single-speaker speech as input, but in addition also the speech mixture.
Here, the separation, mixture and \gls{MAS} encoders consist of 6 Conformer blocks each, all in the same configuration as the baseline blocks.
In total, the \gls{AM} contains 87M parameters without and 127M with mixture encoder.

A LibriSpeech baseline model is first trained for 15 full epochs on the regular LibriSpeech data.
We use it to initialize the \gls{AM} and fine-tune it using the \gls{fCE} loss for up to one epoch on the simulated training data.
The checkpoint is selected based on the best error rate on the dev set.
Longer fine-tuning did not improve the performance.
When using the mixture encoder, the baseline Conformer blocks are split into two parts.
Blocks 1-6 are used to initialize both the separation encoder as well as the mixture encoder while the \gls{MAS} encoder is initialized with blocks 7-12.
The concatenated separation and mixture encoder outputs $\mathbf{h} = [\mathbf{h}_{sep}, \mathbf{h}_{mix}] \in \mathbb{R}^{2N}$ are multiplied with a trainable matrix $\mathbf{W}$ (\enquote{Linear} in \Cref{fig:architecture_mixenc}) to form the \gls{MAS} encoder's input.
For initialization, we set
\begin{equation}
    \mathbf{W} = \begin{bmatrix} \alpha\cdot\text{diag}(\textbf{1}) \\ \beta\cdot\text{diag}(\textbf{1}) \end{bmatrix} + \boldsymbol{\eta} \in \mathbb{R}^{2N\times N}
\end{equation}
so that $\mathbf{h}\cdot\mathbf{W} = \alpha\cdot\mathbf{h}_{sep} + \beta\cdot\mathbf{h}_{mix} + \mathbf{h}\cdot\boldsymbol{\eta}$.
We use $\alpha=0.9$, \\$\beta=0.1$ and uniform random noise in the range of $-0.01$ to $0.01$ for $\boldsymbol{\eta}$.
$N=512$ is the output dimension of the Conformer block.
For some experiments, we additionally apply sequence-discriminative training.
The model is initialized with the \gls{fCE}-fine-tuned parameters and trained further for half an epoch with the lattice-based \gls{sMBR} loss \cite{gibson2006smbr}.
Again, longer training did not improve the performance.

Unlike \cite{berger2023interspeech}, the separator and recognizer are not jointly fine-tuned in this work.
All \glspl{AM} are trained on a single \gls{GPU} using the
\iffinal
RETURNN toolkit \cite{zeyer2018returnn}.
\else
[blind during review] toolkit.
\fi
For the baseline encoder, one full epoch of fine-tuning takes around \SI{17}{\hour} using the \gls{fCE} loss and \SI{120}{\hour} using the \gls{sMBR} loss on an NVIDIA GeForce GTX 1080 Ti.
When adding the mixture encoder, we use an NVIDIA A10 \gls{GPU} and one full epoch takes around \SI{15}{\hour} or \SI{60}{\hour}, respectively.

\begin{table*}[ht]
    \centering
    \makebox[0pt][c]{\parbox{1.\textwidth}{%
        \begin{minipage}[c]{0.28\hsize}
            \vspace{-19.2mm}

\input{tables/results_librispeech}

        \end{minipage}
        \hfill
        \setcounter{table}{2}
        \begin{minipage}[c]{0.70\hsize}

\input{tables/results_literature_overlap}

        \end{minipage}%
        \setcounter{table}{1}
    }}
    \vspace{-3mm}
\end{table*}
\input{tables/results_main}
\setcounter{table}{3}

\vspace{-3mm}
\subsection{Recognition and scoring}
\vspace{-1mm}
\label{sec:recognition_scoring}
For recognition, the separator from \Cref{sec:speech_separator} is applied to separate the \libricss recordings into two overlap-free audio streams.
As mentioned in \Cref{sec:speech_separator}, this still allows processing an arbitrary number of speakers following the \gls{CSS} approach assuming that overlap is limited to two speakers at any time.
A simple energy-based \gls{VAD} is used on the outputs to identify the speech segments.
It uses a threshold and additional smoothing to overestimate the activity which avoids information loss for the \gls{ASR} system \cite{boeddeker2023tssep}.
Subsequently, we recognize the speech with the model from \Cref{sec:acoustic_model} using the
\iffinal
RASR toolkit \cite{rybach2011rasr}.
\else
[blind during review] toolkit.
\fi
Either the official LibriSpeech 4gram or a Transformer \gls{LM} \cite{irie2019trafolm} is used.
The recipes for \gls{AM} training and recognition are available online%
\iffinal
\footnote{\url{https://github.com/rwth-i6/i6_experiments}}.
\else
\footnote{Link blind during review.}.
\fi

As the segments' speaker labels are unknown, it is not possible to compute the regular \gls{WER}.
After obtaining the hypotheses for each \gls{VAD} segment, we concatenate the segments that belong to the same separation output channel.
Finally, the concatenated hypotheses are scored with the MeetEval toolkit\footnote{\url{https://github.com/fgnt/meeteval}} to compute the \gls{ORCWER} \cite{neumann2023wer}.
Future work could consider diarization to compute the \gls{cpWER}, but it is out of scope here.

\section{Results}
\vspace{-1mm}
\subsection{Baseline Results}
\vspace{-2mm}
The results of the clean single-speaker LibriSpeech baseline model on the standard test sets are shown in \Cref{table:results_librispeech}.
Applying this model directly to the \libricss data yields the results marked as no fine-tuning in \Cref{table:results_main}.
It is striking that \tfgridnet outperforms the \gls{BLSTM} separator by a large margin, reducing the \gls{ORCWER} by about a factor of 3.
When the system is fine-tuned on the simulated meeting data, improvements can be observed for both separators.

\vspace{-2mm}
\subsection{Effect of Mixture Encoder}
Adding the mixture encoder further shows a clear improvement for the \gls{BLSTM} separator.
However, for \tfgridnet, the mixture encoder has no advantage over the baseline encoder.
In general, the additional mixture information can only help if the separation is imperfect.
Although \tfgridnet does not provide a perfect separation, it still reaches a point where the effect of correcting separation artifacts using additional mixture information becomes negligible.
This can be seen as a strong supporting argument for the already good performance.
To be fair, it should be noted that, unlike \tfgridnet, the \gls{BLSTM} separator does not use self-attention.
It might therefore benefit more from the combination with the Conformer-based mixture encoder, although it is only slightly worse than a SepFormer model on reverberant data \cite{cord2022monaural}.

\vspace{-2mm}
\subsection{State-of-the-Art and Oracle Results}
In \Cref{table:results_literature_overlap}, we compare the best results from \Cref{table:results_main} to the literature and oracle experiments.
Our results outperform the previous state-of-the-art using only LibriSpeech data \cite{kanda2022tsot} by 26\% relative.
Furthermore, they even outperform \cite{boeddeker2023tssep} which uses a WavLM \cite{chen2022wavlm} model that is 2.5 times larger than our \gls{AM} and was (pre-)trained on about 100 times more data. %
Note that \cite{kanda2022tsot,boeddeker2023tssep} do not use an external \gls{LM}.

In our setup, the \tfgridnet output streams are segmented by a simple \gls{VAD} model (see \Cref{sec:recognition_scoring}).
To assess the limits of this separation, we utilize the boundary times from the \libricss annotation to segment the separated channels.
The output signal for a segment is determined as the separated channel with the maximum \gls{SDR} to the clean LibriSpeech signal of the corresponding segment.
\Cref{table:results_literature_overlap} presents these results as \enquote{ours + oracle segmentation}.
Note that the separator might still output parts of the segment on the other output stream in that case.
As an upper bound, the \enquote{oracle signals and segmentation} performance is obtained by recognizing \textit{test-clean} with the LibriSpeech baseline model and then concatenating and scoring the hypotheses in a way analog to \libricss.
It is also possible to compute the regular \gls{WER} in this case and it is identical to the \gls{ORCWER}.
We can observe that despite its strong performance, the \tfgridnet results are still clearly behind the use of the oracle clean data.
Around a third of the gap can be explained by the oracle segmentation.
The remaining difference is caused by the advantage of oracle separation and enhancement.

\vspace{-2mm}
\subsection{Frame-Wise Error Analysis}

We aim to get a deeper understanding of the influence of separation errors on the recognition performance.
In particular, the goal is to determine how often words from the cross speaker leak to the separated primary channel and end up in the \gls{ASR} search space.
While signal-level metrics struggle to handle silence appropriately and may measure errors that have no effect on the ASR performance \cite{iwamoto22artifacts}, other metrics that operate purely on the transcription (e.g. using Levenshtein distances) do not guarantee temporal locality of detected errors and thus might e.g. blur leakage effects.
In contrast, our proposed analysis based on Hamming distances of frame-wise word-level alignments allows for an accurate attribution of cross channel effects on the ASR search space and 1-best hypothesis.

In particular, we compare forced alignments of the oracle transcription on the primary or cross channel to alignments obtained by computing word lattices or 1-best hypotheses on the primary channel.
For the oracle clean audio we compute the forced alignments with the LibriSpeech baseline \gls{ASR} model and for the separated audio we use our model with \tfgridnet separator and \gls{fCE}-fine-tuned baseline encoder from \Cref{table:results_main}.
The word boundaries obtained from the alignments can be used to assign a word to each frame \cite{wessel2001errorminimization}, where the frame shift is \SI{10}{\milli\second}.
The Hamming distance is then computed on a per-frame basis.
To obtain the \gls{HER}, the distance is divided by the number of frames in the signal.
The \gls{CIR} is the ratio of frames in which both sequences contain the same word.
It is defined as ${\text{CIR} = 1 - \text{HER}}$.
Simply speaking, we measure where the hypothesis did not contain the word from the forced alignment with the \gls{HER}.
Meanwhile, the \gls{CIR} indicates where the primary channel hypothesis contains the same word as the cross channel forced alignment, so potential leakage.
For these frame-wise measures, all frames that contain silence in both sequences are excluded.
\Cref{fig:fer} depicts an example computation.
In contrast, Levenshtein error rates ignore silence entirely and are normalized over the length of the reference transcription.

\input{figures/fer_visualization}

To obtain the cross-talker transcriptions, we derive word boundaries from a forced alignment of the oracle transcription on the primary channel.
Using the segmentation information, we check whether a word falls within the boundaries of a cross channel utterance and assign it to the transcription if applicable.
The resulting transcription contains all words that were uttered by competing speakers and that are fully within the boundaries of the respective utterance.

We use the oracle segmentation and assume that for each segment, the correct separated channel is known as well as that no speaker changes occur within a segment.
For lattice creation, the same bigram \gls{LM} as in the \gls{sMBR} training stage is used.
To compute the Hamming distance between a frame-wise alignment and a word lattice, we check all active word arcs in the lattice at each frame and count an error if the reference word is not among the active hypotheses.
Note that this distance is optimistic since it checks for the best word irrespective of the arc and the path in the lattice (see scoring of \enquote{great} in different paths in \Cref{fig:fer}).

\input{tables/results_fers_ref}

\input{tables/results_fers_cross}

For reference, \Cref{table:results_fers_ref} shows Levenshtein and Hamming error rates between a forced alignment of the oracle transcription and a word lattice as well as a recognized hypothesis on the primary channel of the \libricss test data.
Generally, they are of a similar order of magnitude.
These relatively low error rates demonstrate that our models are generally able to recognize the uttered words correctly and at the correct position in time.

In \Cref{table:results_fers_cross}, we report \glspl{CIR}, i.e. the ratio of frames with equal assignment in both sequences, between the primary and cross channel.
A forced alignment of the oracle transcription on the separated cross channel is compared to a forced alignment of the oracle transcription, a lattice and the 1-best hypothesis on the separated primary channel.
The comparison of both alignments with the oracle transcription shows that in the regions where both speakers are active, the cross-talker utters the same word as the primary speaker in only 0.2\% of the frames on \libricss test.
This can be seen as the natural true coincidence.
When moving to the full search space in the lattice, the \gls{CIR} increases to 0.6\% in overlapped regions,  which now also includes leakage.
This is significantly lower than the overlap on phone level (4.1\%).
The high increase of the \gls{CIR} in regions with one active speaker is mainly caused by frames where only the primary speaker is active but a silence arc is still present in the lattice, e.g. the eighth frame for the \gls{CIR} in \Cref{fig:fer}.
This is reflected when only counting coincidences of real words, where the \gls{CIR} drops from 2.3\% to 0.1\% in \Cref{table:results_fers_cross}.
For the single best hypothesis during recognition, the \gls{CIR} of 0.3\% in regions with overlap is only 0.1\% higher than the actual coincidence measured in the oracle case.
This suggests that the cross-talker is well suppressed and there is hardly any leakage that plays a role in the search space for the primary channel.
Therefore, our results render methods that exploit oracle knowledge about the cross channel transcription during training (e.g. multi-output sequence discriminative training in \cite{zhehuai2018multi_speaker_sequence_training}) in combination with separators of \tfgridnet's quality as unpromising.
Yet, the separation is not perfect as can be seen from the increasing \gls{WER} for higher overlap ratios in \Cref{table:results_literature_overlap}.

\vspace{-1mm}
\section{Conclusion}
\vspace{-1mm}
In this work, we integrate the \tfgridnet separator into a meeting separation and recognition system.
Furthermore, the mixture encoder approach is extended to a setup for an arbitrary number of speakers and transferred to a Conformer-based neural architecture.
Our best system reaches a new state-of-the-art on \libricss and even outperforms a much larger model using additional data.
In addition, we study the effect of exploiting mixture speech encoding for acoustic modeling and show that, unlike previously used separators, \tfgridnet advances the separation quality to a level where the effect of additional mixture information vanishes.
Finally, our analysis demonstrates that the cross-talker is suppressed sufficiently to have only a negligible effect on the primary channel's search space.
The main contributors to the remaining gap to the oracle performance are thus segmentation errors, reverberation as well as non-leakage-related separation artifacts.

\iffinal
\vspace{-1mm}
\section{Acknowledgements}
\vspace{-1mm}
The authors thank Moritz Gunz for help with the LibriSpeech baseline model and Wilfried Michel for helpful discussions on sequence discriminative training.
This research was partially funded by the Deutsche Forschungsgemeinschaft (DFG, German Research Foundation) under project No. 448568305.
Computational Resources were provided by BMBF/NHR/PC2.
\fi

\bibliographystyle{IEEEbib}
\bibliography{refs}

\end{document}

%% file: acronyms.tex
\newacronym{AM}{AM}{acoustic model}
\newacronym{ASR}{ASR}{automatic speech recognition}
\newacronym[longplural={bi-directional long-short term memories}]{BLSTM}{BLSTM}{bi-directional long-short term memory}
\newacronym{BSS}{BSS}{blind speech separation}
\newacronym{CART}{CART}{classification and regression tree}
\newacronym{CE}{CE}{cross entropy}
\newacronym{CER}{CER}{character error rate}
\newacronym{CDp}{CDp}{context dependent phoneme}
\newacronym{CIR}{CIR}{coincidence rate}
\newacronym{CNN}{CNN}{convolutional neural network}
\newacronym{CSS}{CSS}{continuous speech separation}
\newacronym{CTC}{CTC}{connectionist temporal classification}
\newacronym{DNN}{DNN}{deep neural network}
\newacronym{DNN-HMM}{DNN-HMM}{deep neural network hidden Markov model}
\newacronym{FER}{FER}{frame error rate}
\newacronym{fCE}{f-CE}{frame-wise cross-entropy}
\newacronym{GER}{GER}{graph error rate}
\newacronym{GMM}{GMM}{Gaussian mixture model}
\newacronym{GMM-HMM}{GMM-HMM}{Gaussian mixture hidden Markov model}
\newacronym{GPU}{GPU}{graphics processing unit}
\newacronym{HER}{HER}{Hamming error rate}
\newacronym{HMM}{HMM}{hidden Markov model}
\newacronym{LM}{LM}{language model}
\newacronym[longplural={long-short term memories}]{LSTM}{LSTM}{long-short term memory}
\newacronym{MAS}{MAS}{mixture-aware speaker}
\newacronym{MSE}{MSE}{mean squared error}
\newacronym{NN}{NN}{neural network}
\newacronym{PIT}{PIT}{permutation invariant training}
\newacronym{SA-SDR}{SA-SDR}{source-aggregated signal-to-distortion ratio}
\newacronym{SDR}{SDR}{signal-to-distortion ratio}
\newacronym{SNR}{SNR}{signal to noise ratio}
\newacronym{sMBR}{sMBR}{state-level minimum Bayes risk}
\newacronym{STFT}{STFT}{short time Fourier transform}
\newacronym{VAD}{VAD}{voice activity detection}
\newacronym{WER}{WER}{word error rate}
\newacronym{cpWER}{cpWER}{concatenated minimum-permutation \glsentryshort{WER}}
\newacronym{ORCWER}{ORC WER}{optimal reference combination \glsentryshort{WER}}
\newacronym{WERR}{WERR}{word error rate reduction}

%% file: commands.tex
\newcommand{\libricss}{LibriCSS\xspace}
\newcommand{\tfgridnet}{TF-GridNet\xspace}

%% file: figures/architecture.tex
\begin{figure}[!hb]
    \vspace{-4mm}
    \begin{subfigure}[t]{\columnwidth}
        \centering
        \includegraphics[scale=.47]{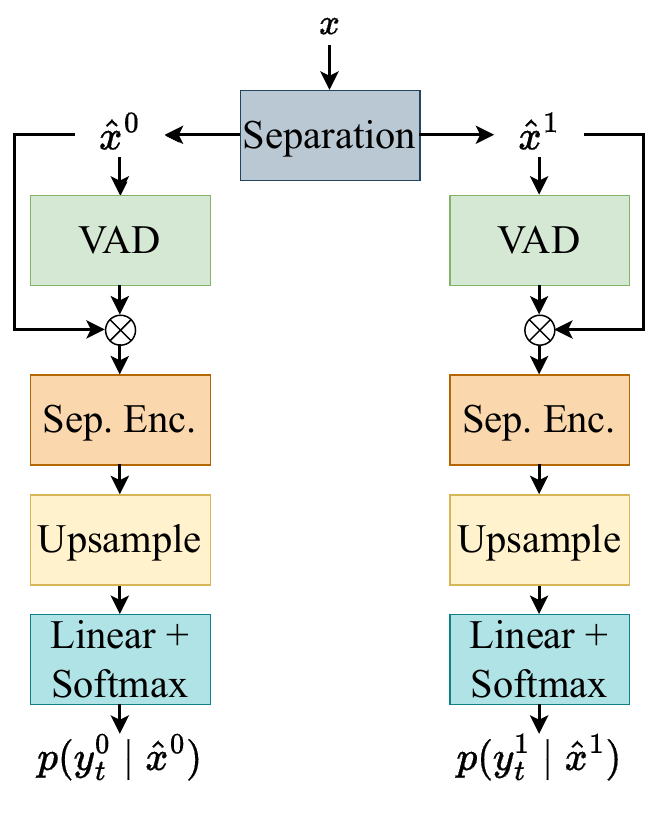}
        \vspace{-4mm}
        \caption{Base architecture.}
        \label{fig:architecture_base}
    \end{subfigure}
    \par
    \begin{subfigure}[b]{\columnwidth}
        \centering
        \includegraphics[scale=.47]{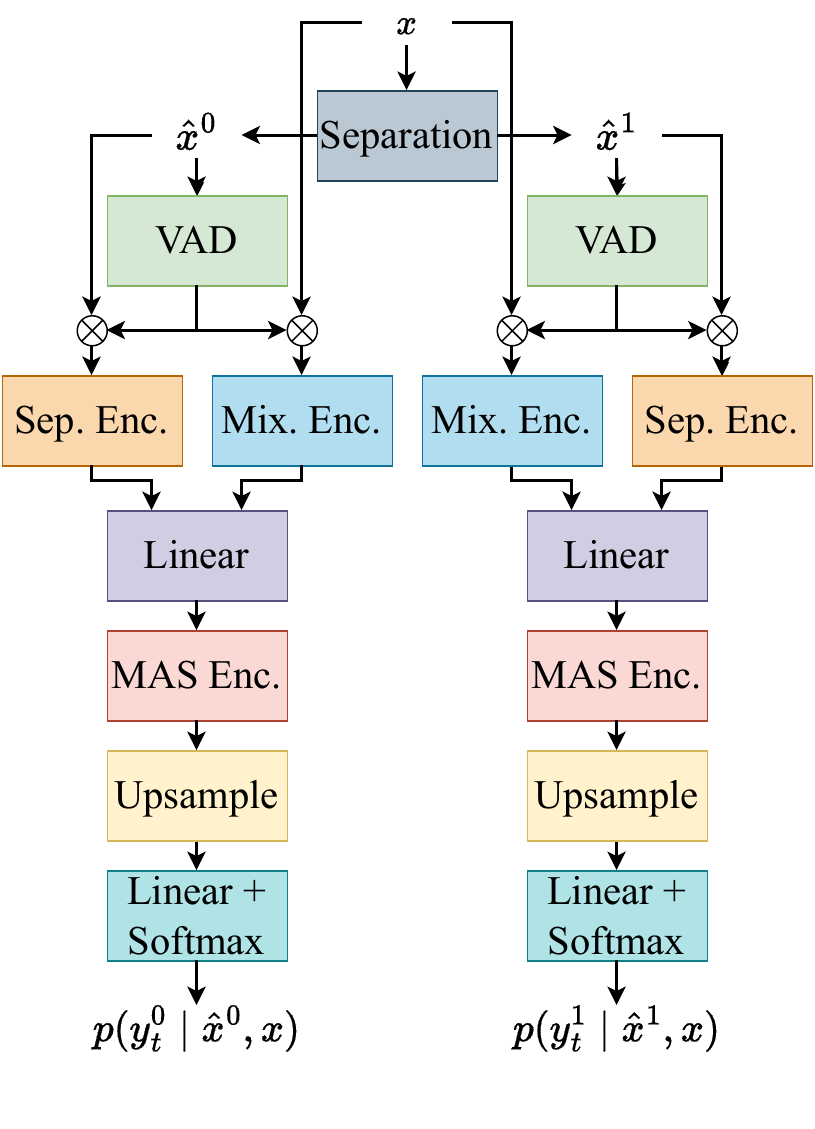}
        \vspace{-4mm}
        \caption{Architecture with mixture encoder.}
        \label{fig:architecture_mixenc}
    \end{subfigure}
    \caption{Model architectures with (a) baseline and (b) mixture encoder for acoustic modeling. Same colors denote shared parameters.}
    \label{fig:architecture}
\end{figure}

%% file: tables/results_librispeech.tex
\centering
\caption{WER [\%] for the LibriSpeech baseline model on the standard LibriSpeech dev and test sets.}\vspace{-2mm}
\label{table:results_librispeech}
\begin{tabular}{|c|c|c|c|c|}
\hline
\multirow{3}{*}{LM} & \multicolumn{4}{c|}{WER [\%]} \\\cline{2-5}
  \multirow{2}{*}{} &      \multicolumn{2}{c|}{dev} & \multicolumn{2}{c|}{test} \\\cline{2-5}
                    &                         clean & other &                     clean & other \\\hline\hline
              4gram &                           2.9 &   6.6 &                       3.2 &   6.8 \\\hline
            Transf. &                           1.9 &   4.3 &                       2.3 &   4.7 \\
\hline
\end{tabular}

%% file: tables/results_literature_overlap.tex
\centering
\caption{WER [\%] on \libricss test compared to other works in the literature. LS960 denotes the LibriSpeech training data, Mix94k denotes the 94k hours WavLM pre-training data. The error rates are roughly comparable but were calculated in different ways: asclite \acs{WER} for t-SOT, \acs{cpWER} with oracle speaker label assignment for TS-SEP, \acs{ORCWER} for ours and oracle.}\vspace{-2mm}
\label{table:results_literature_overlap}
\begin{tabular}{|c|c|c|S|S|S|S|S|S[table-format=2.1]|S|}
\hline
    \multirow{2}{*}{Description} & \multirow{2}{*}{Training Data} & \multirow{2}{*}{LM} & \multicolumn{7}{c|}{WER [\%] for different overlap ratios} \\\cline{4-10}
                                 &                                &                     &                                                       {0S} & {0L} & {10} & {20} & {30} & {40} & {Avg.} \\\hline\hline
      t-SOT \cite{kanda2022tsot} &                          LS960 &  \multirow{2}{*}{-} &                                                        5.3 &  5.4 &  6.5 &  7.3 &  9.5 & 11.3 &    7.6 \\\cline{1-2}\cline{4-10}
TS-SEP \cite{boeddeker2023tssep} &                 Mix94k + LS960 &                     &                                                        3.6 &  3.0 &  4.0 &  5.9 &  8.0 &  8.5 &    5.8 \\\hline&&&&&&&&&\\[-2.2ex]\hline
                Ours (\tfgridnet &         \multirow{6}{*}{LS960} &               4gram &                                                        5.5 &  4.3 &  5.0 &  6.5 &  8.5 & 10.0 &    6.9 \\\cline{3-10}
                  and mix. enc.) &                                &             Transf. &                                                        4.5 &  3.5 &  3.6 &  5.2 &  7.1 &  8.4 &    5.6 \\\cline{1-1}\cline{3-10}&&&&&&&&&\\[-2.2ex]\cline{1-1}\cline{3-10}
                   Ours + oracle &                                &               4gram &                                                        4.1 &  3.7 &  4.0 &  5.1 &  6.9 &  8.4 &    5.6 \\\cline{3-10}
                    segmentation &                                &             Transf. &                                                        2.9 &  2.7 &  2.7 &  3.5 &  5.4 &  6.6 &    4.1 \\\cline{1-1}\cline{3-10}&&&&&&&&&\\[-2.2ex]\cline{1-1}\cline{3-10}
                  Oracle signals &                                &               4gram &                                                        3.1 &  3.2 &  3.0 &  3.1 &  3.2 &  3.1 &    3.1 \\\cline{3-10}
                and segmentation &                                &             Transf. &                                                        2.2 &  2.4 &  1.9 &  2.0 &  2.1 &  2.0 &    2.1 \\
\hline
\end{tabular}

%% file: tables/results_main.tex
\begin{table}[h]

\centering
\caption{ORC-WER [\%] on \libricss test (average over all overlap ratios) for different combinations of separators and \acs{AM} encoders. In all cases, the model is trained on clean LibriSpeech first. If fine-tuning is applied, the model is further fine-tuned on simulated meeting data separated by the respective separator.}\vspace{-2mm}
\label{table:results_main}
\begin{tabular}{|c|c|c|c|S|}
\hline
                   Separator &                Fine-Tuning &                   Encoder &      LM & {\acs{ORCWER} [\%]} \\\hline\hline
      \multirow{6}{*}{BLSTM} &      \multirow{2}{*}{None} & \multirow{4}{*}{baseline} &   4gram &                21.6 \\\cline{4-5}
                             &                            &                           & Transf. &                19.4 \\\cline{2-2}\cline{4-5}
                             & \multirow{4}{*}{\acs{fCE}} &                           &   4gram &                19.9 \\\cline{4-5}
                             &                            &                           & Transf. &                17.9 \\\cline{3-5}
                             &                            &  \multirow{2}{*}{mixture} &   4gram &                19.3 \\\cline{4-5}
                             &                            &                           & Transf. &                17.6 \\\hline
                             &      \multirow{2}{*}{None} & \multirow{6}{*}{baseline} &   4gram &                 7.9 \\\cline{4-5}
                             &                            &                           & Transf. &                 6.4 \\\cline{2-2}\cline{4-5}
                             & \multirow{2}{*}{\acs{fCE}} &                           &   4gram &                 7.2 \\\cline{4-5}
                             &                            &                           & Transf. &                 5.8 \\\cline{2-2}\cline{4-5}
TF-                          &                \acs{fCE} + &                           &   4gram &                 7.0 \\\cline{4-5}
GridNet                      &                 \acs{sMBR} &                           & Transf. &                 {\textbf{5.6}} \\\cline{2-5}
                             & \multirow{2}{*}{\acs{fCE}} &  \multirow{4}{*}{mixture} &   4gram &                 7.2 \\\cline{4-5}
                             &                            &                           & Transf. &                 5.8 \\\cline{2-2}\cline{4-5}
                             &                \acs{fCE} + &                           &   4gram &                 6.9 \\\cline{4-5}
                             &                 \acs{sMBR} &                           & Transf. &                 {\textbf{5.6}} \\
\hline
\end{tabular}\vspace{-4mm}

\end{table}

%% file: figures/fer_visualization.tex
\begin{figure}[tb]
    \includegraphics[width=0.95\columnwidth]{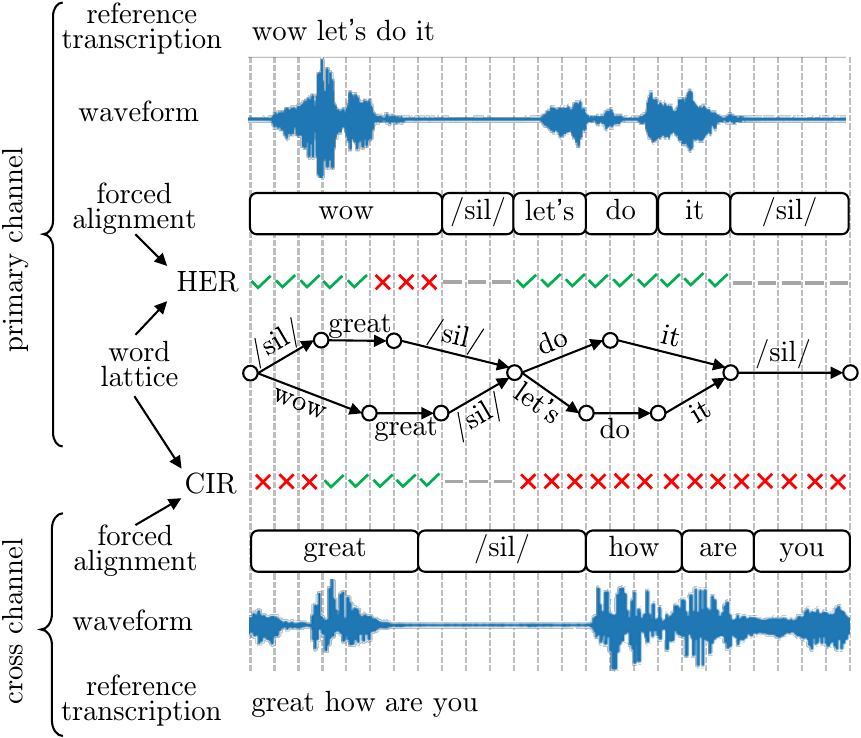}
    \caption{
        Visualization of the frame-wise error computation.
        The vertical gray lines represent frame boundaries.
        Frame-wise error measures are drawn for a comparison between primary channel forced alignment and primary channel word lattice as well as a comparison between primary channel word lattice and cross channel forced alignment.
        We indicate whether the word labels match ({\color{green}\ding{52}}), mismatch ({\color{red}\ding{54}}) or both channels' forced alignments contain silence ({\color{gray}\textbf{\Large{--}}}).
        The upper comparison within the primary channel corresponds to the row with separated audio and lattice hypothesis in \Cref{table:results_fers_ref} and in this example, we obtain $\text{HER} =3/17\approx 18\%$.
        The lower comparison between primary and cross channel corresponds to the row with lattice hypothesis for the primary channel in \Cref{table:results_fers_cross} and the example results in $\text{CIR} =5/22\approx 23\%$.
        The frames are not drawn to scale.        
    }
    \label{fig:fer}
    \vspace{-3mm}
\end{figure}

%% file: tables/results_fers_ref.tex
\begin{table}[tb]

\centering
\caption{Error rates [\%] between a forced alignment of the oracle transcription vs. lattice/1-best hypothesis on \libricss test. The audio is either the oracle clean audio or the separated primary channel. The Levenshtein error rate corresponds to the \gls{GER} for the lattice and to the regular \gls{WER} for the 1-best hypothesis.}
\label{table:results_fers_ref}
\begin{tabular}{|c|c|c|c|c|}
\hline
       Audio & Hypothesis & \acs{LM} & \multicolumn{2}{c|}{Error rate [\%]} \\\cline{4-5}
             &            &          &                          Levenshtein & Hamming \\\hline\hline
Oracle clean &    Lattice &   bigram &                                  1.4 &     3.5 \\\cline{2-5}
             &     1-best &    4gram &                                  3.1 &     3.8 \\\cline{3-5}
             &            &  Transf. &                                  2.1 &     3.3 \\\hline
   Separated &    Lattice &   bigram &                                  3.2 &     3.0 \\\cline{2-5}
             &     1-best &    4gram &                                  5.9 &     5.5 \\\cline{3-5}
             &            &  Transf. &                                  4.4 &     4.4 \\
\hline
\end{tabular}
\vspace{-3mm}
\end{table}

%% file: tables/results_fers_cross.tex
\begin{table}[tb]

\centering
\caption{Analysis of the cross speaker's influence on the primary speaker's channel via leakage. We report \glspl{CIR} [\%], i.e., the ratio of frames in which both sequences contain the same word, between a forced alignment of the oracle transcription on the separated cross channel and a forced alignment of the oracle transcription, a lattice and the 1-best hypothesis on the separated primary channel. Coincidences are counted once including silence and once for real words only. Results on \libricss test.}
\label{table:results_fers_cross}
\begin{tabular}{|c|c|c|S|S|S|S|S|S|}
\hline
   Hyp. & \acs{LM} & Level & \multicolumn{6}{c|}{Coincidence rate [\%]} \\\cline{4-9}
Primary &          &       &        \multicolumn{3}{c|}{Words and sil.} &      \multicolumn{3}{c|}{Words only} \\\cline{4-9}
        &          &       &       \multicolumn{3}{c|}{\#act. speakers} & \multicolumn{3}{c|}{\#act. speakers} \\\cline{4-9}
        &          &       &                                        {1} & {2} & {Avg.} &                                  {1} & {2} & {Avg.} \\\hline\hline
 Oracle &      N/A & Phone &                                        0.0 & 4.1 &    0.5 &                \multicolumn{3}{c|}{} \\\cline{3-9}
        &          &  Word &                                        0.0 & 0.2 &    0.0 &                                  0.0 & 0.2 &    0.0 \\\cline{1-2}\cline{4-9}
Lattice &   bigram &       &                                        2.3 & 0.6 &    2.1 &                                  0.1 & 0.6 &    0.1 \\\cline{1-2}\cline{4-9}
 1-best &    4gram &       &                                        0.2 & 0.3 &    0.2 &                                  0.0 & 0.3 &    0.1 \\\cline{2-2}\cline{4-9}
        &  Transf. &       &                                        0.4 & 0.3 &    0.4 &                                  0.0 & 0.3 &    0.1 \\
\hline
\end{tabular}
\vspace{-4mm}
\end{table}